\documentstyle[12pt,A4]{article}
\newcommand{\bi}{\bibitem}

\begin{document}
\baselineskip = 18pt

\begin{center}
{\LARGE\bf
Full determination of the $dp$
backward elastic scattering matrix element.}\\
\vspace{1cm}
{\Large
V~P~Ladygin
and N~B~Ladygina$\footnote{E-mail: ladygina@sunhe.jinr.dubna.su}$}
\vspace{0.5cm}
~\\
{\noindent  LHE-JINR, 141980, Dubna, Moscow Region, Russia } \\
\end{center}
\vspace{1.5cm}

\baselineskip = 18pt

\begin{quote}

{\large\bf Abstract.}
The model-independent analysis of the $dp$ elastic scattering in the collinear geometry
has been performed.
It is shown that the measurements of 10 polarization observables of the first
and second order realize the complete experimental program on the determination of
the amplitudes of the $dp$ backward elastic scattering reaction.

\end{quote}

\vspace{1cm}
{\large\bf PACs: 24.70.+s, 25.45.-z}
\vspace{0.3cm}

\newpage

\section{Introduction}
~

The structure of the deuteron is extensively studied at last
decades using both electromagnetic and hadron probes.
This interest is related with the hope to explore the internal deuteron
structure over a wide range of distances between constituents.
The  backward elastic scattering $dp\to pd$ at medium
and high energies is one of the simplest processes with
the large momentum transfer and, therefore, can be used to
study the high-momentum tail of the deuteron wave function (DWF).
Another interesting feature of this process is that within
framework of the one nucleon exchange (ONE) the cross section is
proportional to the fourth degree of the DWF, $|\Phi(q)|^4$,
and polarization observables are also simply related with the
$S$- and $D$- components of the DWF.

The experimental data on the differential cross section
of the $dp\to pd$ reaction \cite{crsec}
show a sharp peak at $180^o$ in the center of mass.
%The simplest mechanism that leads to this peak is ONE.
On the other hand, the differential
cross section at $\theta\sim 180^o$
demonstrates the strong energy dependence and  an enhancement
in vicinity of the $\Delta$-isobar excitation.
This resonant-like energy dependence of the
cross section could not be
explained by the pole mechanism only.
In the model of Kerman and Kisslinger \cite{kerman}
the $dp$ backward elastic peak is interpreted as due to the
admixture of $NN^*(1688)$ state in the standard deuteron wave
function.
The cross section of the $dp\to pd$ process was calculated in the framework of
the two-step model in which the cross section of the $dp$ backward elastic
scattering is expressed in
terms of the  $pp\to d\pi^+$ cross section.
Such mechanism, considered firstly by Craigie and Wilkin \cite{wilkin}
and later by Barry \cite{barry}
leads to a resonant behaviour of the cross section at energies near
$T_p\sim 600~MeV$.
Calculations of Kolybasov and Smorodinskaya \cite{kolyb}
taking into account $D$- state and relativistic corrections are
in the satisfactory agreement with the cross section data.

%which has a maximum in  the region of
%the $\Delta$ production in the intermediate state.
%Craigie and Wilkin \cite{wilkin} as well as Barry \cite{barry} explain
%this enhancement as an effect
%of pion exchange.
%They proposed models where $pd\to dp$ cross section can be
%expressed in terms of $pp\to d\pi^+$ cross section.
%Craigie and Wilkin calculated the cross section of
%$pd$ backward elastic scattering.
%In this model the $dp\to pd$ cross section is proportional
%to the cross section of the reaction $pp\to d\pi^+$,
%which has a maximum in the indicated energy region because
%of the $\Delta$ production in the intermediate state.

In the model developed by
Kondratyuk and Lev \cite{lev},
the $dp\to pd$ reaction amplitude is expressed directly in terms of
the $NN\to N\Delta$ amplitudes.
It is
shown that the interference of the $\Delta$ excitation with the
pole mechanism
could provide a satisfactory description of the energy dependence of
the $dp$ backward elastic scattering cross section.

The description of the cross section data improves significantly when the contribution from
the three-baryon resonances obtained in the bag model as the nine-quark states
with hidden color are added \cite{3bar}.

However, description of the polarization observables is related
with the numerous difficulties.
The two-step model predicts the simple relations between
polarization observables of the $dp$ backward elastic scattering
and the $pp\to d\pi^+$ process, which should coincide.
Moreover, the measurements of analyzing power due to
the polarization of incident proton have shown that
these relations do not hold even in vicinity of
the $\Delta$ resonance.
Measurements of the vector analyzing power $A_y$ for
$\vec{p}d\to dp$ at large angles at 316 and 516 $MeV$ and
comparison with the $pp\to d\pi^+$  data have shown that this
analyzing power is higher than that for the
corresponding $\vec{p}p\to d\pi^+$ reaction \cite{ander}.

Measurements of the tensor analyzing power
$T_{20}$ performed at Saturne \cite{arv1} at  $T_d\sim  0.3\div
2.3~ GeV$ shown the large negative value for $T_{20}$.
Calculations of Boudard and Dillig \cite{boudard} taking into account
ONE and rescattering mechanisms from the
$\Delta$ resonance fails to reproduce the behavior of
$T_{20}$.
The better agreement with the experimental data
was obtained by Nakamura and Satta \cite{japon} in the framework of
the two-step model by imposing of T-invariance on the triangle amplitudes.
The results have shown the strong sensitivity to the results of
the $pp\to d\pi^+$ phase shift analysis and possible existing of
dibaryon resonances at $\sqrt{s} \sim 2.1\div 2.2~GeV$.
Recent measurements of $T_{20}$ and spin transfer
coefficient from vectorially polarized deuteron to
the proton, $\kappa_o$, performed at Saclay \cite{exp249} have confirm
the deviation from  ONE model.
Measurements of $T_{20}$ at Dubna up to  $T_d\sim 5~GeV$
have demonstrated the new unexplained to date structure at
$T_d \sim 3\div 3.2~GeV$ \cite{exp250}.

%The goal of present article is to find out the set of
%observables to provide the full determination of the matrix element
%of $dp$ backward elastic scattering reaction.
In present article we perform the model-independent analysis of
the $dp$ backward elastic scattering reaction in the collinear
geometry.
The purpose of this analysis is to find  the minimal and optimal
set of the experiments for the direct reconstruction of the
reaction amplitudes.
In the next section we derive the matrix element of the
$dp$ elastic scattering.
% in the collinear kinematics.
In section 3 we give the expressions for the number of polarization
observables. In section 4 we find the set of observables to reconstruct
the amplitudes of the reaction.
In the last section we discuss the experimental possibilities to
perform this set of measurements.

%\newpage

\section{$dp$ elastic scattering in the collinear geometry}
~

In the general case the
$1+\frac{1}{2}\to 1+\frac{1}{2}$ process can be described in
terms of 18 independent complex amplitudes:
\begin{eqnarray}
\label{matrix0}
& & {\cal M} = \chi^+_f{\cal F}\chi_i,\nonumber\\
& & {\cal F} = f_1 (\vec{\xi_1}\vec{l})(\vec{\xi_2}^+\vec{l}) +
f_2 (\vec{\xi_1}\vec{m})(\vec{\xi_2}^+\vec{m})+
f_3 (\vec{\xi_1}\vec{n})(\vec{\xi_2}^+\vec{n})+
f_4 (\vec{n}~\vec{\xi_1}\times\vec{\xi_2}^+)\nonumber\\
& &+f_5 \big ((\vec{\xi_1}\vec{l})(\vec{\xi_2}^+\vec{m})+
(\vec{\xi_1}\vec{m})(\vec{\xi_2}^+\vec{l})\big )+
(f_6 (\vec{\xi_1}\vec{l})(\vec{\xi_2}^+\vec{l})+
f_7 (\vec{\xi_1}\vec{m})(\vec{\xi_2}^+\vec{m})\\
& & +f_8 (\vec{\xi_1}\vec{n})(\vec{\xi_2}^+\vec{n})) (\vec{\sigma}\vec{n})+
(f_9 (\vec{m}~\vec{\xi_1}\times\vec{\xi_2}^+)+
f_{10} (\vec{l}~\vec{\xi_1}\times\vec{\xi_2}^+))(\vec{\sigma}\vec{l})+
(f_{11} (\vec{m}~\vec{\xi_1}\times\vec{\xi_2}^+)\nonumber\\
& & +f_{12} (\vec{l}~\vec{\xi_1}\times\vec{\xi_2}^+))(\vec{\sigma}\vec{m})+
f_{13} (\vec{n}~\vec{\xi_1}\times\vec{\xi_2}^+)(\vec{\sigma}\vec{n})+
f_{14} ((\vec{\xi_1}\vec{l})(\vec{\xi_2}^+\vec{n}) +
(\vec{\xi_1}\vec{n})(\vec{\xi_2}^+\vec{l}))(\vec{\sigma}\vec{l})\nonumber\\
& & +f_{15} ((\vec{\xi_1}\vec{m})(\vec{\xi_2}^+\vec{n}) +
(\vec{\xi_1}\vec{n})(\vec{\xi_2}^+\vec{m}))(\vec{\sigma}\vec{l})+
f_{16} ((\vec{\xi_1}\vec{l})(\vec{\xi_2}^+\vec{n}) +
(\vec{\xi_1}\vec{n})(\vec{\xi_2}^+\vec{l}))(\vec{\sigma}\vec{m})\nonumber\\
& & +f_{17} ((\vec{\xi_1}\vec{m})(\vec{\xi_2}^+\vec{n})
(\vec{\xi_1}\vec{n})(\vec{\xi_2}^+\vec{m}))(\vec{\sigma}\vec{m})+
f_{18} ((\vec{\xi_1}\vec{l})(\vec{\xi_2}^+\vec{m}) +
(\vec{\xi_1}\vec{m})(\vec{\xi_2}^+\vec{l}))(\vec{\sigma}\vec{n}),\nonumber
\end{eqnarray}
where $\chi_f$ and  $\chi_i$ are the spinors of the final and initial fermions,
$\vec{\xi_1}$ and $\vec{\xi_2}$ are
the polarization vectors of the initial and final spin 1 particles;
$f_{\it i}$ are scalar amplitudes, depending in the general case on the
energy and scattering angle $\theta$.
The three mutually orthogonal unit vectors $\vec{l}$, $\vec{m}$ and $\vec{n}$
are defined as:
\begin{eqnarray}
\vec{l}= \frac{\vec{k_i} + \vec{k_f}}{|\vec{k_i} + \vec{k_f}|},~~~~~
\vec{m}= \frac{\vec{k_i} - \vec{k_f}}{|\vec{k_i} - \vec{k_f}|},~~~~~
\vec{n}= \frac{\vec{k_i}\times \vec{k_f}}{|\vec{k_i}\times \vec{k_f}|},
\end{eqnarray}
where $\vec{k_i}$ and $\vec{k_f}$ are the relative momenta of the initial and
final states, respectively.

The expression (\ref{matrix0}) have been obtained using the parity conservation
only. Therefore, it keeps the terms violating the T-invariance in
the $dp$ elastic scattering. The requirement of time reversal invariance
implies that:
\begin{eqnarray}
\label{TRI}
f_5 = f_9 = f_{12} = f_{15} = f_{16} = f_{18} = 0,
\end{eqnarray}
 what reduces the number of independent amplitudes for $dp$ - elastic scattering
to 12.

In this paper we consider the $dp$ elastic scattering in the collinear
geometry.
%assuming T-invariance can be violated.
% f_{16}=-f_{18}
The condition of the total spin projection conservation for the collinear processes restricts the  number of independent amplitudes
of the matrix element  (\ref{matrix0}) in such a way that:
\begin{eqnarray}
& & f_2=f_3,~~~f_{11}=f_{13},\nonumber\\
& & f_4=f_6=f_7=f_8=f_{14}=f_{17}=0
\end{eqnarray}
for the forward elastic scattering ($\theta_{cm} = 0^o$) and
\begin{eqnarray}
& &f_1=f_3,~~~f_{10}=f_{13},\nonumber\\
& &f_4=f_6=f_7=f_8=f_{14}=f_{17}=0
\end{eqnarray}
for the backward elastic scattering ($\theta_{cm} = 180^o$), respectively.

Finally, the amplitude of the $dp$ elastic scattering in the collinear
geometry can be written as:
\begin{eqnarray}
\label{matrix}
{\cal F} = & & A(\vec{\xi_1}\vec{\xi_2}^+) + B(\vec{\xi_1}\vec{k}) (\vec{\xi_2}^+\vec{k}) +{\it i} C(\vec{\sigma}\vec{\xi_1}\times \vec{\xi_2}^+) +
{\it i} D (\vec{\sigma}\vec{k})(\vec{k}\vec{\xi_1}\times \vec{\xi_2}^+),
\end{eqnarray}
where $\vec{k}$ is the unit vector in the direction of
the relative momentum of the initial state, $A$, $B$, $C$ and $D$ are
the amplitudes
of the  $dp$ elastic scattering depending on the energy.

%Note, that forward and backward $dp$ elastic scattering have the same spin %structure, however,
%the amplitudes $A-D$ are different in both cases.

\section{Polarization observables}
~

In this section we give the definition of the general polarization observable
%in terms of density matrices of involved particles
and results of our calculations for the number of polarization observables
of the $dp$ elastic scattering in the collinear kinematics.

%\subsection{ General polarization observable}

%The polarized cross section can be written in terms of the
%initial density max
We define the general spin observable in
terms of the Pauli $2\times 2$ spin matrices $\sigma$ for protons and a set of
spin operators $S$ for deuterons as in refs\cite{igo1,igo2,alberi}:
\begin{eqnarray}
\label{general}
C_{\alpha,\lambda,\beta,\gamma}=
\frac{Tr({\cal F}{\sigma_\alpha}{S_\lambda}{\cal F^+}{\sigma_\beta}{S_\gamma})}{Tr({\cal F}{\cal F^+})},
\end{eqnarray}
where indices $\alpha$ and $\lambda$ refer to the initial proton and
deuteron polarization, indices $\beta$ and $\gamma$ refer to the final proton
and deuteron, respectively;  $\sigma_0$ and $S_0$ corresponding to the
non-polarized particles are the unit matrices of two and
tree dimensions in these notations.
We use a righthand coordinate system,
defined in accordance with Madison convention \cite{madison}. This system is specified by a
set of three ortogonal vectors $\vec{L}$, $\vec{N}$ and  $\vec{S}$,
where  $\vec{N}=\vec{n}$, $\vec{L}=\vec{l}$  and $\vec{S}=[\vec{N}\vec{L}]$.

Below we shall show that the direct reconstruction of the amplitudes
can be provided by the measurements of the first and second order
polarization observables only, therefore, we derive the expressions only
for them.

%\subsection{Cross section and analyzing powers}

%In this section we derive the expressions for
%polarization observables using the matrix element (\ref{matrix}).
The squared matrix element (\ref{matrix}) is expressed as:
\begin{eqnarray}
\label{f2}
Tr{\cal{FF}}^+ = 2 (3A^2+ 2{\cal R}eAB^* + B^2 + 6C^2 + 4 {\cal R}eCD^* + 2D^2)
\end{eqnarray}

The only non-vanishing polarization observable of the first order is
the tensor analyzing power
due to the polarization of the initial deuteron
(or the induced tensor polarization of the final deuteron) can be written as:
\begin{eqnarray}
\label{t20}
Tr{\cal {FF}}^+ C_{0,NN,0,0}= 2\big ( 2{\cal R}eAB^* +B^2 -
2{\cal R}e CD^*- D^2 \big )
\end{eqnarray}
This observable
$T_{20} = -\sqrt{2}\cdot  C_{0,NN,0,0}$ was measured up to $5~GeV$ of the initial deuteron kinetic energy at Saclay and Dubna
\cite{arv1,exp249,exp250}.

The tensor polarization of
the final deuteron is equal to the tensor analyzing power due to polarization of the initial deuteron according to the T-invariance:
\begin{eqnarray}
\label{r20}
C_{0,0,0,NN} = C_{0,NN,0,0}
\end{eqnarray}

%\subsection{ Deuteron-deuteron spin transfer coefficients}

The spin transfer coefficients from deuteron to
deuteron due to the vector polarization of
both particles can be expressed as:
\begin{eqnarray}
\label{sptrvvd}
Tr{\cal {FF}}^+C_{0,L,0,L} & = & 4\cdot (A^2 + C^2 + 2{\cal R}e CD^* +D^2)\\
Tr{\cal {FF}}^+C_{0,N,0,N} & = & 4\cdot (A^2 + {\cal R}e AB^* + C^2)
\end{eqnarray}

Since the deuteron is a spin 1 particle, there are the number of
tensor-tensor and vector-tensor (tensor-vector) spin transfer coefficients
apart from the usual vector-vector ones.
The tensor-tensor  non-vanishing spin transfer coefficients are defined
as:
\begin{eqnarray}
& &Tr{\cal {FF}}^+ C_{0,LL,0,LL}=4\big (3A^2 + 4{\cal R}eAB^* + 2B^2 - 3C^2 + 2{\cal R}e CD^* + D^2 \big )~~~~~~~\\
& &Tr{\cal {FF}}^+ C_{0,NN,0,NN}=2\big (6 A^2 + 2{\cal R}eAB^* + B^2 - 6C^2 -8 {\cal R}e CD^* -4 D^2 \big )\\
& &Tr{\cal {FF}}^+ C_{0,NN,0,SS}=2(-3A^2+2{\cal R}eAB^*+B^2+3C^2+10{\cal R}eCD^*+5D^2)\\
& &Tr{\cal {FF}}^+ C_{0,LN,0,LN}={9}\cdot\big ( A^2 + {\cal R}eAB^* -C^2 \big )\\
& &Tr{\cal {FF}}^+ C_{0,SN,0,SN}={9}\cdot\big ( A^2  -C^2  -2{\cal R}eCD^* - D^2\big)
\end{eqnarray}
Note that tensor-tensor spin transfer coefficients
$C_{0,NN,0,NN}$, $C_{0,NN,0,SS}$, $C_{0,NN,0,LL}$, $C_{0,LL,0,LL}$, and
$C_{0,SN,0,SN}$ are not independent and related  as the
following:
\begin{eqnarray}
\label{ttrelation}
C_{0,NN,0,LL} & = & -\frac{1}{2} C_{0,LL,0,LL}\nonumber\\
C_{0,SN,0,SN} & = & \frac{1}{2} (C_{0,NN,0,NN} - C_{0,NN,0,SS})\\
C_{0,LL,0,LL} & = & {2} ( C_{0,NN,0,NN} + C_{0,NN,0,SS})\nonumber
\end{eqnarray}
The existence of these relationships allows to perform the
measurements of $C_{0,NN,0,NN}$ and $C_{0,NN,0,SS}$ instead
$C_{0,SN,0,SN}$ and $C_{0,LL,0,LL}$, being more difficult to be
realized  from the experimental point of view.

The spin transfer coefficient from the vectorially polarized deuteron to
the tensorially polarized final deuteron depends on the $A$ and $B$ amplitudes
only:
\begin{eqnarray}
\label{sptrvtd}
Tr{\cal {FF}}^+C_{0,N,0,LS}  & = & 6\cdot {\cal I}mAB^*
\end{eqnarray}
There are the simple relations between
the tensor-vector
%spin transfer coefficients from the tensorially polarized deuteron to the
%deuteron vectorially polarized
and vector-tensor spin transfer coefficients due to the T-invariance and
rotation symmetry:
\begin{eqnarray}
\label{sptrtvd}
C_{0,LS,0,N} = -C_{0,LN,0,S}= - C_{0,N,0,LS} = C_{0,S,0,LN}
\end{eqnarray}

%\subsection { Spin correlations }

The availability of the polarized proton target and polarized beam gives the
opportunity to measure the different spin correlations.
As in case of the spin transfer coefficients from the deuteron to deuteron,
there are the spin correlations due to the vector polarization of the deuteron:
\begin{eqnarray}
\label{spcovvd}
Tr{\cal {FF}}^+C_{L,L,0,0} & = & 4\cdot (2{\cal R}eAC^* + 2{\cal R}eAD^* - C^2)\\
Tr{\cal {FF}}^+C_{N,N,0,0} & = & 4\cdot (2{\cal R}eAC^* + {\cal R}eBC^* -{\cal R}eCD^* - C^2),
\end{eqnarray}
as well as due to tensor polarization:
\begin{eqnarray}
\label{spcovvd1}
Tr{\cal {FF}}^+C_{N,LS,0,0}  =  6\cdot  ({\cal I}m (B + D)C^*)
\end{eqnarray}
The measurement of the spin correlations of the final particles is
not realistic due to the small cross section of the considered process.
Moreover, since
they are equal to the
corresponding spin correlations of the initial particles
according to the T-invariance:
\begin{eqnarray}
\label{spcovvf}
C_{0,0,L,L} & = & C_{L,L,0,0}~~~~~~~~~
C_{0,0,N,N}  =  C_{N,N,0,0}\nonumber\\
C_{0,0,S,S} & = & C_{0,0,N,N}~~~~~~~~~
C_{0,0,N,LS}  =  C_{N,LS,0,0},
\end{eqnarray}
these experiments are not necessary.
%because they do not provide the additional information.

%\subsection{ Proton-proton spin transfer coefficients}
%\subsection{ Deuteron-proton spin transfer coefficients}

The  deuteron-proton
spin transfer coefficients can be written as:
\begin{eqnarray}
\label{sptrdp}
Tr{\cal {FF}}^+C_{0,L,L,0} & = & 4\cdot (2{\cal R}eAC^* + 2{\cal R}eAD^* + C^2)\\
Tr{\cal {FF}}^+C_{0,N,N,0} & = & 4\cdot ({\cal R}e(2A^* +B^* + D^*)C + C^2)\\
Tr{\cal {FF}}^+C_{0,LS,N,0} & = & 6\cdot ( {\cal I}m(-B^* + D^*)C)
\end{eqnarray}
Only $C_{0,N,N,0} = \frac{2}{3}\kappa_0$ was measured at Saturne to date \cite{exp249}.

Finally, we give the  expressions for
the spin transfer coefficients from polarized proton target to the final proton:
\begin{eqnarray}
\label{sptrvvp}
Tr{\cal {FF}}^+C_{L,0,L,0} & = & 2\big (3A^2 + 2{\cal R}eAB^* + B^2 - 2C^2\big )\\
Tr{\cal {FF}}^+C_{N,0,N,0} & = & 2\big (3A^2 + 2{\cal R}eAB^* + B^2 - 2C^2 - 4{\cal R}eCD^*- 2D^2 \big )\\
C_{S,0,S,0} & =& C_{N,0,N,0}
\end{eqnarray}
In this section we have derived the expressions for all non-vanishing observables
of the first and second order, which can be used to determine the matrix element
of the process.
Some of these observables are
not independent due to the symmetries properties. As the consequence
of the T-invariance one can write the relationships between
the different observables:
\begin{eqnarray}
\label{tinv}
C_{0,0,0,NN}  &=& C_{0,NN,0,0}~~~~~~C_{0,NN,0,SS} = C_{0,SS,0,NN}
~~~~~~C_{0,LL,0,NN} = C_{0,NN,0,LL}\nonumber\\
C_{L,0,0,L} & = & C_{0,L,L,0}~~~~~~~~
C_{N,0,0,N}  =  C_{0,N,N,0}~~~~~~~~
C_{N,0,0,LS}  =  C_{0,LS,N,0}
\end{eqnarray}
There are  also the following relations
due to the collinear geometry of the considered process:
\begin{eqnarray}
\label{coll}
C_{S,S,0,0} & =& C_{N,N,0,0}~~~~~~~~
C_{0,S,0,S}   =  C_{0,N,0,N}~~~~~~~~
C_{S,0,S,0}   =  C_{N,0,N,0}\nonumber\\
C_{0,0,S,S} & =& C_{0,0,N,N}~~~~~~~~
C_{0,LN,0,LN}  =  C_{0,LS,0,LS}
\end{eqnarray}

%To conclude this section, we have derived the expressions for
%all non-vanishing observables of the first and second order
%for the $dp$ elastic scattering in the collinear geometry.

%\newpage

\section{Direct reconstruction of the amplitudes}
~

Since the $dp\to pd$ process is described by 4 complex amplitudes,
one needs to measure at least 7 observables to determine the matrix
element. On the other hand, all polarization observables are
expressed through the bilinear combinations of the amplitudes,
therefore, the number of experiments  to be performed at given energy
increases.
Another restriction comes from small cross section of
the process at high energies and low efficiency of the polarimeters or
limited possible flux on the polarized target.
Therefore, one needs to find  the set of observables satisfying
to the experimental possibility to be measure also.

Below we assume that the axis of the primary deuteron beam polarization at the exit of the ion source is vertical.

The moduli of the amplitudes $A$, $B$ and $C$ can be extracted from
the information on the cross section, tensor analyzing power
$C_{0,NN,0,0}$ (or $C_{0,0,0,NN}$) and 3 different tensor-tensor spin transfer
coefficients: $C_{0,NN,0,LL}$, $C_{0,SN,0,SN}$ (see relations (\ref{ttrelation})) and  $C_{0,LN,0,LN}$ as:
\begin{eqnarray}
A^2 &=& \frac{1}{18} Tr{\cal{FF}}^+\big ( 1 - \frac{1}{2} C_{0,NN,0,LL} -
2~C_{0,NN,0,0} +C_{0,SN,0,SN}\big )\\
B^2 &=& -\frac{1}{4} Tr{\cal{FF}}^+ \big ( C_{0,NN,0,LL} + \frac{2}{9}
(4C_{0,LN,0,LN} -C_{0,SN,0,SN})\big )\\
C^2 &=& \frac{1}{18} Tr{\cal{FF}}^+ \big (1 + C_{0,NN,0,LL} +
C_{0,NN,0,0}\big )
\end{eqnarray}
Note, that $C^2$ can be also reconstructed from the
vector-vector,  $C_{0,N,0,N}$, and tensor-tensor,   $C_{0,LN,0,LN}$,
spin transfer coefficients from the deuteron to deuteron:
\begin{eqnarray}
C^2 = \frac{1}{8} Tr{\cal{FF}}^+ \big ( C_{0,N,0,N} -\frac{4}{9}C_{0,LN,0,LN}\big ).
\end{eqnarray}
In such a way $C_{0,N,0,N}$ is not independent and can be expressed through the
discussed above observables as:
\begin{eqnarray}
C_{0,N,0,N} = \frac{4}{9}\big ( 1 + C_{0,NN,0,LL} + C_{0,NN,0,0} +
C_{0,LN,0,LN}\big )
\end{eqnarray}
%it is not necessary to measure this observable.

The reconstruction of the $D^2$ requires the measurement of the additional
observables because the $D^2$ appears always
together with the $2{\cal R}eCD^*$ in
the expressions for the cross section, tensor analyzing power and
tensor-tensor spin transfer coefficients, and therefore, can
not be extracted.
One can easy to show that the $D^2$  can be obtained from the 3 polarization
observables: spin transfer coefficient from deuteron to proton,
$C_{0,N,N,0}$ \cite{exp249}, spin correlation parameter, $C_{N,N,0,0}$ \cite{slr},
and spin transfer coefficient from the proton to proton, $C_{N,0,N,0}$:
\begin{eqnarray}
D^2 = \frac{1}{4} Tr{\cal{FF}}^+\big ( \frac{1}{2} \big ( 1 - C_{N,0,N,0} ) + C_{N,N,0,0}
- C_{0,N,N,0}\big )\big )
\end{eqnarray}
On the other hand, the spin transfer coefficient $C_{N,0,N,0}$ is not independent
observable and related with the tensor analyzing power $C_{0,NN,0,0}$ and
tensor-tensor spin transfer coefficient $C_{0,NN,0,NN}$ by:
\begin{eqnarray}
C_{N,0,N,0} = \frac{1}{9} \big ( 1 + 4~C_{0,NN,0,0} + 4~C_{0,NN,0,NN}\big )
\end{eqnarray}
Therefore, $D^2$ can be obtained by the measuring of the spin correlation $C_{N,N,0,0}$ and
spin transfer coefficient from the deuteron to proton, $C_{0,N,N,0}$ \cite{exp249}, in
addition to the cross section, tensor analyzing power and tensor-tensor
spin transfer coefficient:
\begin{eqnarray}
D^2 =\frac{1}{9}Tr{\cal{FF}}^+ \big( 1 - \frac{1}{2}(C_{0,NN,0,0} + C_{0,NN,0,NN})
+ \frac{9}{4} ( C_{N,N,0,0} - C_{0,N,N,0})\big )
\end{eqnarray}

The reconstruction of the phases of the amplitudes  can be performed using
an additional information apart from the discussed above observables.
%requires the measurement of additional observables.
Since the observables are expressed through the bilinear combinations
of the amplitudes, the reconstruction of the common phase is impossible.
We put the phase of the amplitude $A$ equal to  zero, $\Phi_A = 0$.
Therefore $A$ amplitude is real:
\begin{eqnarray}
{\cal I}mA = 0 ~~~~~~~~~ {\cal R}eA =\sqrt{A^2} = |A|
\end{eqnarray}

Imaginary part of the $B$ amplitude can be easy reconstructed
from the measurement of the spin transfer coefficient from
the vectorially polarized initial deuteron to the tensorially polarized final
deuteron $C_{0,N,0,LS}$. Therefore, the phase $\Phi_B$ can be reconstructed as:
\begin{eqnarray}
{\cal I}mB & =& |B|~\sin{\Phi_B}=\frac{1}{6~|A|}Tr{\cal {FF}}^+ C_{0,N,0,LS}\nonumber\\
{\cal R}eB & =& |B|~\cos{\Phi_B}=
\frac{1}{|A|}\big (\frac{1}{9} Tr{\cal {FF}}^+ C_{0,LN,0,LN} - A^2 + C^2\big )
\end{eqnarray}

The reconstruction of the ${\cal R}eC$ and ${\cal I}mC$ can be obtained from
the spin correlation $C_{N,LS,0,0}$ and  spin
transfer coefficient  $C_{0,LS,N,0}$ due to the tensor polarization of the
initial deuteron and polarization of the initial and final proton,
respectively.
Using the following combinations of the polarization observables
\begin{eqnarray}
X & =& \frac{1}{12} Tr{\cal {FF}}^+ \big ( C_{N,LS,0,0} + C_{0,LS,N,0} \big ) = {\cal I}mBC^*\nonumber\\
Y & =& \frac{1}{8} Tr{\cal {FF}}^+ \big ( C_{N,N,0,0} + C_{0,N,N,0} \big ) =
2~{\cal R}eAC^* + {\cal R}eBC^*\nonumber,
\end{eqnarray}
one can reconstruct the phase $\Phi_C$:
\begin{eqnarray}
{\cal I}mC & =& |C|~\sin{\Phi_C}=\frac{Y~{\cal I}mB - X~(2~|A|+{\cal R}eB)}{2~|A|{\cal R}eB + B^2}\nonumber\\
{\cal R}eC & =& |C|~\cos{\Phi_C}=\frac{X~{\cal I}mB + Y~{\cal R}eB}{2~|A|{\cal R}eB + B^2}
\end{eqnarray}

The reconstruction of the real and imaginary parts of the $D$ amplitude
can be performed without
measurements of the additional observables.
Using the following relations:
\begin{eqnarray}
U & =& -\frac{1}{12} Tr{\cal {FF}}^+ \big ( C_{N,LS,0,0} + C_{0,LS,N,0} \big ) = {\cal I}mCD^*\nonumber\\
V & =& \frac{1}{8} Tr{\cal {FF}}^+ \big ( C_{N,N,0,0} - C_{0,N,N,0} \big ) -C^2 =
{\cal R}eCD^*\nonumber,
\end{eqnarray}
one can easy obtain $\Phi_D$:
\begin{eqnarray}
{\cal I}mD &=& |D|~\sin{\Phi_D}= \frac{1}{C^2}\big ( V {\cal I}mC - U {\cal R}eC \big )\nonumber\\
{\cal R}eD &=& |D|~\cos{\Phi_D}= \frac{1}{C^2}\big ( V {\cal R}eC + U {\cal I}mC \big )
\end{eqnarray}

It should be noted, that the matrix element can be expressed in terms
of different sets of amplitudes  which  are simply related (see Appendix 1).
The minimal number of different experiments (seven) does not depend on
the used set of amplitudes.
The expressions of the amplitudes $A-D$ calculated within One Nucleon
Exchange are given in Appendix 2.

%\newpage

\section {Conclusions and discussions}
~

To conclude we have found that the measurement of
10 observables of the first and second order only, i.e.
cross section, tensor analyzing power $C_{0,NN,0,0}$ ($C_{0,0,0,NN}$)
and 8 second order spin observables $C_{0,NN,0,NN}$,
$C_{0,NN,0,SS}$, $C_{0,LN,0,LN}$, $C_{0,N,0,LS}$, $C_{0,N,N,0}$,
$C_{N,N,0,0}$, $C_{LS,N,0,0}$ and $C_{0,LS,N,0}$
could provide the direct reconstruction of the amplitudes of
the $dp$ backward elastic process. The chosen observables are mostly realistic
to be measured at the moment with the existing experimental techniques.

Using of purely tensor polarimeter POLDER \cite{polder} based on
the $(d,2p)$ charge exchange reaction \cite{wilkin1} could provide
the measurements of $C_{0,NN,0,NN}$, $C_{0,NN,0,SS}$ and $C_{0,LN,0,LN}$ (which
could be easy obtained by the rotation of the initial deuteron spin) and
$C_{0,N,0,LS}$ in addition to the measured cross section \cite{crsec}
and $C_{0,NN,0,0}$ \cite{arv1,exp249,exp250}.
The cross section and tensor analyzing power could
be measured simultaneously with the spin transfer coefficients
allowed by the triggering system of POLDER \cite{polder}.
Note, that as a by-product
the tensor polarization of the final deuteron $C_{0,0,0,NN}$ also would be obtained.
Two additional observables:
spin transfer coefficients $C_{0,N,N,0}$ \cite{exp249} and $C_{0,LS,N,0}$
could  be obtained from the measurement of the proton polarization by the
polarimeter POMME \cite{pomme}.
The spin correlations $C_{N,N,0,0}$ \cite{slr} and $C_{LS,N,0,0}$ could be
measured using polarized deuteron beam and polarized proton target installed
now at LHE of JINR \cite{lehar}.
The rotation of the primary deuteron spin could be
provided by the magnetic field of the  beam line upstream the target
\footnote{ We thank I.M.Sitnik for the estimation of
the deuteron spin rotation angle due to the beam line upstream
polarized proton target installed at LHE JINR}
or by the special spin-flip magnet, what
is necessary for the measurement of T-odd observables like
$C_{N,LS,0,0}$ and $C_{0,LS,N,0,0}$ or tensor-tensor spin transfer coefficient
$C_{0,LN,0,LN}$.

The measurements of the  tensor-tensor spin transfer coefficients is the most difficult task
from the experimental point of view because of two reasons:
unfavourable jacobian in the laboratory for the final deuteron and small
figures of merit $F_{20}$, $F_{22}$ and $F_{21}$ of POLDER \cite{polder} which  are about
$1\%$. These factors could be compensated by the using of the
full intensity  of the deuteron beam reaching $2\cdot 10^{11}$
per beam burst and a large solid angle.
The measurement of the proton polarization can be performed with the better
precision because of the figure of merit of POMME as a proton polarimeter at
high energies is about $4\div 5\%$ and favourable jacobian
for the final proton, on the other hand.
But an additional problem due to the large yield of the background breakup process
near the backward elastic peak arises \cite{exp250}.
%in comparison with the detection of the deuteron.
The precision of the measurement of the observables requiring the using of the
polarized target is also limited by the maximal possible flux of the charged particles
as $10^8$ per second.

Therefore, to realize the program of the full experiment, it is necessary to
use 2 different setups: high resolution spectrometer with the proton polarimeter
and proton polarized target for the fast final proton and
a large solid angle spectrometer and low energy deuteron polarimeter like
POLDER \cite{polder} or AHEAD \cite{ahead} to measure the polarization
of the final deuteron.
The first setup can be used to measure $C_{N,N,0,0}$, $C_{N,LS,0,0}$, $C_{0,N,N,0}$
and $C_{0,LS,N,0}$, the second one to measure deuteron-deuteron spin transfer coefficients.
Both setups can be used to measure cross section, tensor analyzing power and spin correlations.

It has been shown in the previous section that, for instance, the measurement of the
vector-vector spin transfer coefficient from deuteron to deuteron,
$C_{0,N,0,N}$, or from proton to proton, $C_{N,0,N,0}$,
do not provide an additional information in comparison with the
tensor-tensor spin transfer coefficients. But the using of low energy
deuteron  AHEAD-like polarimeter \cite{ahead}
which has a non-negligiable vector figure of merit
instead purely tensor polarimeter POLDER could provide the simultaneous
measurement of $C_{0,N,0,N}$ or $C_{0,LN,0,S}$.
Some of additional observables (for instance, $C_{0,S,S,0}$) could be measured as
a by-product during the  experiments requiring the rotation of the polarization axis of
tensorially polarized beam which has an admixture of the vector polarization.
This additional information could be used in order to reduce the
systematic errors of the amplitudes reconstruction.
\vspace{1cm}

{\Large\bf Acknowledgements}
\vspace{0.5cm}

The authors express their thanks to
N.M.Piskunov, I.M.Sitnik and M.P.Rekalo for their attention to this work.
The authors are grateful to the members of POLDER team, especially to
C.Furget, S.Kox and J.-S.Real, for useful discussions.
One of authors (VPL) thanks E.Tomasi-Gustafsson for organization of his
staying and the kind hospitality of Laboratoire National
Saturne, where part of this paper was written.

\vspace{1cm}

\newpage

{\Large\bf Appendix 1}
\vspace{0.5cm}

We give here the relations between different sets of amplitudes.

The  helicity amplitudes, $H_{\frac{1}{2}1\to\frac{1}{2}1}$, can be expressed
as following:
\begin{eqnarray}
\label{helicity}
H_{++\to ++} &=& A + C + D\nonumber\\
H_{+0\to +0} &=& A + B\nonumber\\
H_{+-\to +-} &=& A - C -D\nonumber\\
H_{+0\to -+} &=& \sqrt{2}C\nonumber
\end{eqnarray}

The ukrainian amplitudes \cite{slr} are simply related with the amplitudes
used in this work as:
\begin{eqnarray}
\label{ukranian}
g_1 &=& A  \nonumber\\
g_2 &=& A + B   \nonumber\\
g_3 &=& C \nonumber\\
g_4 &=& C + D   \nonumber
\end{eqnarray}

\vspace{0.5cm}

{\Large\bf Appendix 2}
\vspace{0.5cm}
%\appendix

Here we derive the expressions  for $A-D$ amplitudes
calculated in the framework of the pole mechanism:
\begin{eqnarray}
\label{one}
A &=& \left ( u + \frac{w}{\sqrt{2}} \right )^2\nonumber\\
B &=& -\frac{3}{2} w \left ( 2\sqrt{2} u - w \right )\nonumber\\
C &=& \left ( u + \frac{w}{\sqrt{2}} \right ) ( u - \sqrt{2} w )\nonumber\\
D &=& \frac{3}{\sqrt{2}} w \left ( u + \frac{w}{\sqrt{2}} \right ),\nonumber
\end{eqnarray}
where $u$ and $w$ are the
$S$- and $D$- components
of the deuteron wave function.
\vspace{1cm}

\newpage
%\end{document}

\end{document}